# GOptimaEmbed: A Smart SMS-SQL Database Management System for Low-Cost Microcontrollers

**N.E. Osegi**
Department of Information and Communication Technology
National Open University of Nigeria
Lagos State, Nigeria
E-mail: geeqwam@gmail.com
Website: www.osegi.com
Tel: +234 7030081615

**P.Enyindah**
Department of Computer Science
University of Port Harcourt
Rivers State, Nigeria
E-mail: probi2k3@yahoo.com
Tel: +234 8036710489

**ABSTRACT**

The era of the Internet of things (IOT's), machine-to-machine (m2m) and human-to-machine (h2m) computing has heralded the development of a modern-day smart industry in which humanoids can co-operate, co-exist and interact seamlessly. Currently, there are many projects in this area of smart communication and thus giving rise to an industry electrified by smart things. In this paper we present a novel smart database management system (dbms), *GOptimaEmbed*, for intelligent querying of databases in device constrained embedded systems. The system uses genetic algorithms as main search engine and simplifies the query process using stored in-memory model based on an invented device dependent Short-messaging-Structured Query Language, (SMS-SQL) schema translator. In addition, querying is done over the air using integrated GSM module in the smart space. The system has been applied to querying a plant database and results were quite satisfactory.

## 1. INTRODUCTION

A smart dbms is one in which information storage and retrieval is done entirely by smart devices. Smart devices on the other hand are ubiquitous knowledge based devices with inferential abilities. Smart processing involves the intelligent operations, timely intervention, extraction and utilisation of abstract and real data in smart space or environment. The capability of a smart system is thus dependent on the quality and specifications of the smart devices employed which is largely influenced by the nature of the application and the environment. Typical smart devices used include from high-end and high cost microprocessors such as ARM Cortex M3, AVR 32-bit, and System-on-chips (SoC's) to low-end and cheap microprocessors typically of the 8-bit category.

Smart databases are useful when traditional databases may not be a reliable and flexible option as in real-time sensor-actuated information networks and dynamic databases. When the database is not very large and memory requirements are small, a smart dbms may be preferable. Currently, improvement in smart device manufacture and design miniaturization has led to the available of very cheap high performance low-end or resource constrained microprocessor systems such as the Arduino [1]. Smart dbms systems can encourage and facilitate the end-user programming modelling.

Smart systems generally refer to a class of expert system devices for which the level of abstraction is a variant of the input to the system. In smart solutions, two core requirements are essential:

i. The knowledge or information and
ii. The Inferential Mechanism

In order to effectively manipulate such a system, there is need for the presence of at least two main players. We may refer to the actors or players in a smart solution as "humanoids". Here humanoids" is an acronym that stands for machine-man co-operation or human-robot interface systems.

Currently, smart systems employ high-end microprocessors and embedded systems such as

ARM, FPGA, AMD and advanced computation libraries. These architectures may employ a subset of SQL called tinySQL [2] or SQL lite as well as a client-server engine for database management [3]. However, little work has been carried out by researchers on the potential of SMS-SQL database integration in low-end devices. The lack of tight integration in existing systems has also reduced the effectiveness of information retrieval from these systems. This is due to lack of suitable and readily available SMS-SQL syntax translator for in-memory database and an adaptable and robust search algorithm for devices of the low-end category. Since this devices have a useful role to play in the smart world, dbms strategies have to be devised to tackle this category of embedded processors. *GOptimaEmbed*, is an alternative smart dbms using the strength of a genetic algorithmic search model and an efficient SMS-SQL syntax translator for in-memory database queries. The core components of this system will be described briefly here to enable the reader gain an insight into the "world of smart intelligence".

## 1.1 Genetic Algorithms

Genetic algorithms (GA) introduced by Holland [4], are bio-inspired artificial (or machine) learning search procedures developed to find exact or near solutions to a variety of optimality and knowledge discovery problems [5]. The GA's are actually computer programs that reside on a microcomputer device and offer the benefits of bio-inspired artificial reproduction which include random mutations, cross-over, selection and recombination from a parent population, as well as fitness measures and checks. The artificial reproduction is mathematically modelled using natural heuristics algorithms and decision trees with a stopping function when fitness is met. Fitness is actually described as a core mathematical function or logic operation that must be met for an individual to be deemed successful. Artificial reproduction gives rise to offspring's who later become parents themselves after a successful survival (fitting) exercise. These reproduction and fitting process is carried out over many generations. In practice the generations are typically set to a finite number say 10 to 100 for embedded systems and small to medium-scale AI based software projects. The fitted individuals are successful and are said to be potential solutions to the problem. The population in a GA process is made up of individuals also called chromosomes (genes) who participate in the evolutionary process. All the genes in a given population give rise to the gene pool. The Genotype describes the individuals' structure and quality. For a genotype, a defining length is given which is typically of fixed length strings or integer values representation and from which a phenotype may be decoded – the decoding process is practically done using special array structure handling. The phenotype contains the alleles (or gene values). They serve as input values that must be fed to the fitness function to validate the evolution process. The gene locus or location of a gene on a chromosome will be dependant to a large extent on the nature of the cross-over scheme and to a small extent on the type of mutation. Several mutation and crossover schemes have been given in [6] and have been successfully applied in practice. In a time dependent manner a selection is made using any of the selection schemes of roulette wheel, tournament or truncation described in [6, 7, 8]. The advantages of a GA approach is that they can serve as very good approximate reasoners and converge to the local minima quickly and effectively. Thus, through the concept of natural evolution- mutations, crossover, recombination, a GA optimized solution can possibly attain the solution state earlier than conventional systems.

Some GA's for low end microprocessors have been developed in [5, and 9]. However, these algorithms are needlessly complex, and not flexible for database integration. In this regard we seek to implement an effective but simple algorithm on a small footprint microcontroller using a modified GA optimisation scheme adapted from [6]. This has the advantage of a minimum barebones approach to the Genetic search problem and can easily be integrated into a C++ library class for further flexibility.

## 1.2 Statement of Problem

In recent times there has been a rise in dedicated ubiquitous smart microcomputer products for information retrieval and storage. However, most of these products are either too expensive or are not sensor and human friendly. They lack interactive and controllability features. By using high-end databases, they are generally non-customizable making them resemble their PC counterparts. Also, the potential of remote query exploitation using the SMS-SQL approach has not been fully explored in the area of device constrained smart computers.

## 1.3 Research Objectives

Our research objectives are two-fold.
First, we will build a smart computer model that will implement a real-time smart dbms for device constrained devices.
Secondly, we will validate the model using the SMS-SQL queries as a proof-of-concept

### 1.4 Embedded Processors/Systems

Embedded processors are typically microprocessors with a single CPU that emphasises controllability rather than complex computational arithmetic. However, modern day embedded processors can carry our some level of complex arithmetic with some reduction in performance. In some cases these processors are referred to as microcontrollers or single-board-computers. Examples of such include the 8051's, AVR series, ARM Cortex, Arduino, Intel Galileo, and variants thereof. Embedded processors form the heart and soul of an embedded system and provide the foundation for intelligence building in a smart product. In our prototype, we have used the Arduino microcomputer which is based on the ATmega328P, an AVR 8-bit core microprocessor as a representative processor.

### 1.5 GSM Modules

GSM modules are communication devices that come with a full instruction set for interfacing embedded processors. These devices come in various sizes and shapes and allow for remote messaging, internet and voice services via a GSM network. Some typical examples include the Quectel M10 GSM Module, SM5100B Module and the GSM click. We have utilized the Quectel M10 GSM Module since this feature an onboard antennae and integrates the Arduino shield concept.

### 1.6 Device SQL and In-Memory Model Databases

Device SQL refers to SQL for an embedded processor database. This database is often of the Array structure and builds on the internal memory of the device in question. The benefits of Device SQL cannot be overemphasized and has been described in [10]. In designing a Device SQL framework one needs not bother about implementation language, technology or the issue of database drivers. The only main drawback of such systems is that most low-end microprocessors have limited memory constraints and speed of processing. However, this can be worked around by modifying the data structures with strict types. Since the focus is on micro-data and not Big Data, this shouldn't be a problem. Currently modern low cost microprocessors that offer high memory in megabytes range which can support both processing high volumes of image and text are available. With the DeviceSQL concept, the embedded processor should be responsible for its own dbms requirements and only call upon external memory when its internal memory is not sufficient.

In this paper we emphasise on device dependent database since our storage requirements is small and when the need to adapt the system to respond to dynamic inputs. We call this small data "micro-data". This typically takes a few kilobytes and is desirable particularly in Sensor actuated networks.

## 2. RELATED WORKS

Very little work has been done in the area of smart database query SMS-SQL integrating Artificial Intelligence schemes like GA's. In [11] an SMS-SQL has been proposed for the management and querying of a database. The features of the existing system include a high-end network processor and an Operating System (OS) based on the Unix kernel in a client-server environment. The system require additional component such as USB Modem and two ports and this can increase the overall cost of the system. Using a PC-based client-server dependent protocol – see Figure 1, the system still requires traditional database connectivity for its operation. In addition the choice of non-native language (they used Perl language) means increases in cost of memory allocated. Their system thereby suffers from the database connectivity (dbc) bottleneck.

**MGuide:**

This is a System for mobile information access and is based on Java 2 Microedition (j2me) standard [12]. It supports both static and dynamic information access. The system requires two j2me compliant cell-phones for operation. One of the phones acts as a server connected to the internet through PC. The other is the user requesting for information. As stated earlier, MGuide use j2me which has the draw back of lack of support for retrieval of messages from inbox directly from the screen. Another drawback of the system is the use of traditional database which cannot be implemented on low end microcontrollers.

**TinyDB:**

TinyDB [13] is a fully functional SQL equivalent small footprint open-source database for sensor actuated networks using battery-powered nodes or motes (sensor microprocessors with integrated wireless radio). It has been successfully deployed in low-end micro-computing platforms. It typically runs on motes from the Crossbrow Technologies and compiles using the nesC compiler [14]. However TinyDB does not accommodate external queries using SMS since it does not encourage the humanoid model. Nevertheless, it possesses key features that can make low-end microprocessors a success story. Its query syntax can easily be transformed and integrated in both embedded personal computing platforms and standalone systems.

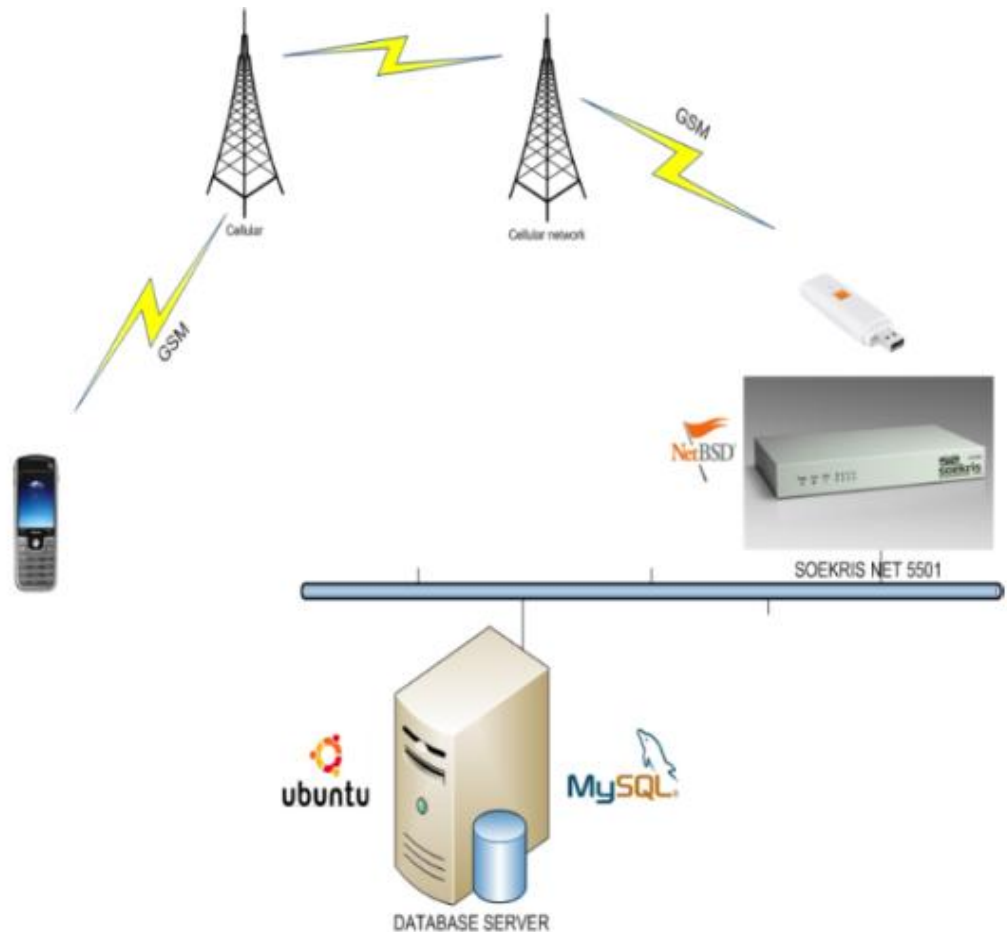

Fig 1: Protocol of an Existing System (Courtesy: Sbaa, 2012)

**ExtremeDB:**

ExtremeDB [15] is a commercial dbms that also support small footprint database for embedded processors with in-memory database features as well as large PC database. It runs on embedded real time operating systems (RTOS), as well as traditional operating systems like windows and linux. No support for SMS-SQL translation has been explicitly defined in their model.

To implement a low-end microprocessor based SMS-SQL dbms we need to employ simple data structures with emphasis on simplified device SQL query definitions, SMS information processing and improved search algorithm such as genetic algorithm. This can be made possible using the *GOptimaEmbed*, framework.

## 3. PROPOSED SYSTEMS MODELLING AND DESIGN

One major challenging in devising an SMS-SQL translator is decoding the required message in microprocessor dynamically, parsing message to data handlers and making the report in a timely and efficient manner. In the design of our systems model we envisage two use cases

1. Static In-memory database (SID) Use-Case

2. Dynamic (Sensor) Actuated In-memory database (DID)

    The *GOptimaEmbed* model follows from the need to realize real-time data acquisition solutions that will facilitate information retrieval from small data repositories (or libraries) as well as real time data from production stations. Since it is hardware as well as software system, it conforms the class of computing referred to as "Physical Computing" [16]. Our software project/modelling methodology shall encourage the functional-object-oriented paradigm with an emphasis on rapid prototyping. We shall focus on this aspect of software engineering based on Physical Computers.

    a. **Systems Protocol**

    The protocol of the proposed system is shown in Figure 2. The sms is transmitted over GSM network to remote server. The server queries the specified database for requested information and reports back to the user the extracted information.

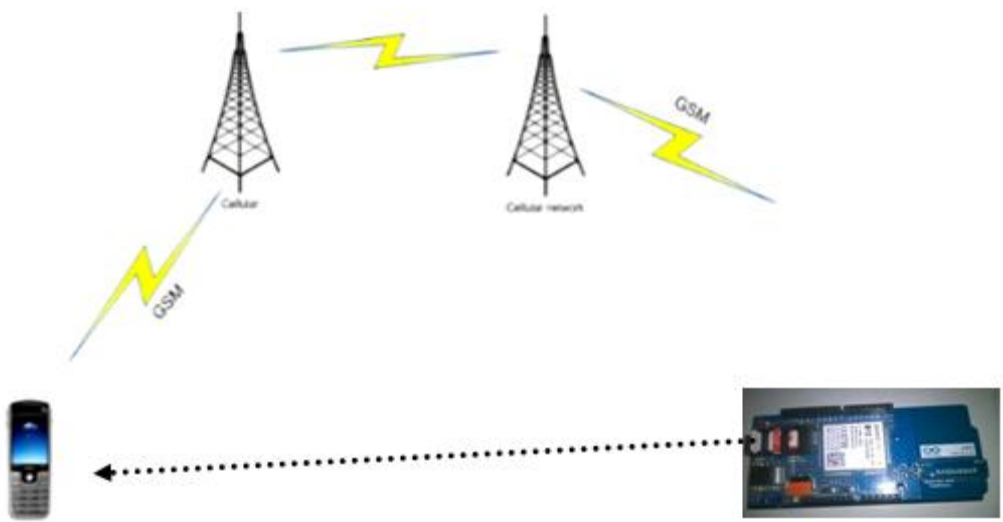

Fig 2: Systems Protocol of Smart dbms

b. **Systems Architecture**

The architecture of *GOptimaEmbed* describes a modular component system with parallel processing channels for simultaneous query retrieval. Concurrency is achieved using multi-tasked time schedule function blocks. To minimize energy consumption the system activates power down mode during inactivity. The system is useful as queries can be redirected to as many repositories as possible using microprocessor firmware. The system is characterized by an applications development framework consisting of GSM class libraries, GSM driver, Network Interfaces, the hardware components, a GA Optimizer and the DeviceSQL In-memory Translator functions. An Arduino Bootloader is included as this provides the enabling platform and mechanism for the system to be implemented, operate and function without necessary any specialized source code downloader hardware [1]. The systems component diagram is shown in Figure 3 below.

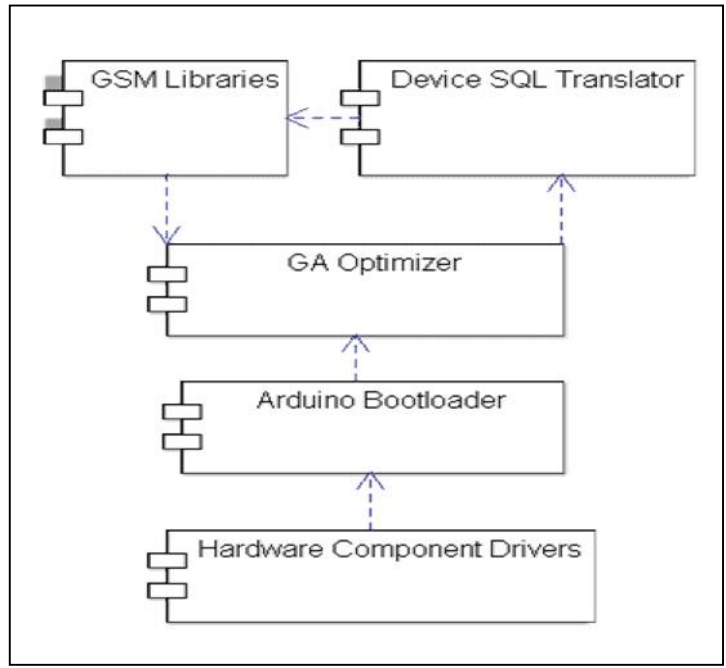

Fig 3: *GOptimaEmbed* Component Architecture

### c. SMS-SQL Query Process Model
The SMS-SQL Query Process Model describes the message format for requesting information from a resource. In Figure 4 is depicted the sequences of states that must be overcome for a smart dbms access in a finite time. This process can generally occur in two stages:
i. Querying Structure
ii. Query Translation and Parser

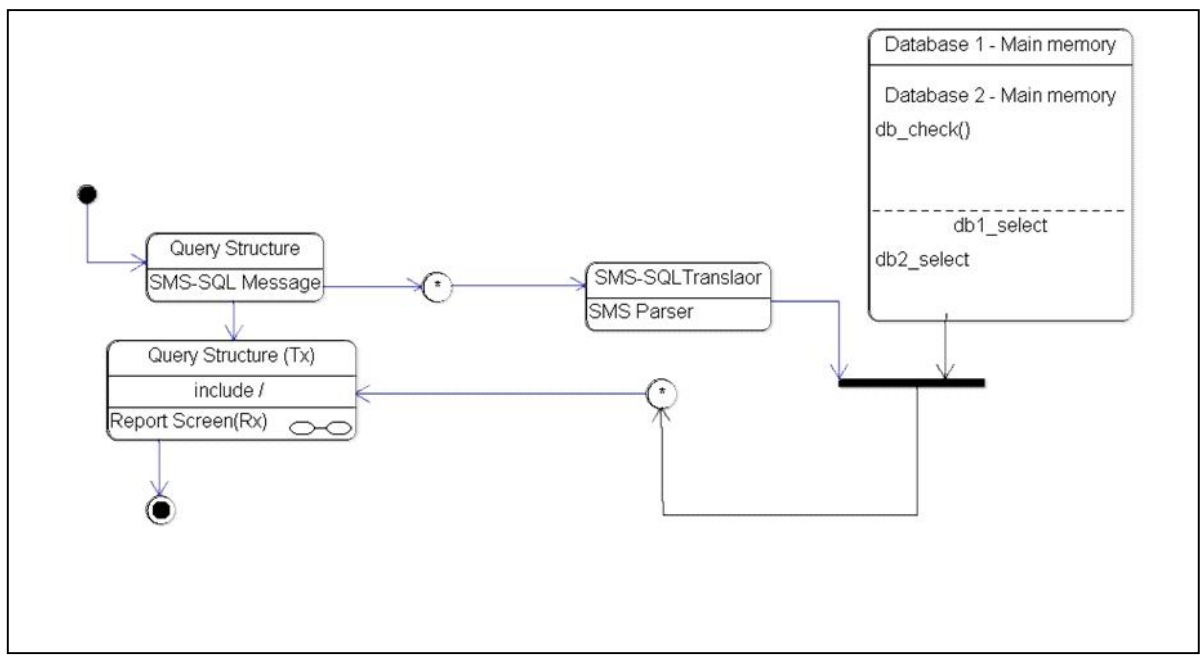

Fig 4: Query Process Model

i. **SMS-**
ii. **SQL Query structure**
Users can send query via sms using simple syntax given as:
*Query: db1 tb1 at1 va4*
Where,
db1 stands for database 1
tb1 stands from table 1
at1 stands for first attribute and,
va4 stands for attribute value of 4.
The SQL equivalent is given as:
*Select Tb1 from Db1 where At1 is = "4"*
This SQL restructuring is necessary to facilitate the end user query entries and to simplify the translation at the Server end.

iii. **SQL- translation**
The translation process involves decoding the necessary parameters from the received query string and discarding irrelevant details. This is done by employing suitable string indexing functions and equivalency commands. The Boolean logic of if-else decision is then used to extract message tokens from the requisite data stores in-memory. This translation process is supported by a genetically optimized search routine the modifications of which is given in Appendix I.

d. **Design**
The design will typically include the specifications, a block level description of the system and the algorithmics. The system is based on the Arduino single-board microcomputer, a powerful and cost-effective prototyping platform for resource constrained microprocessor systems. Smart communication by the humanoids is effected by using the shield concept where a smart communication module is pin-to-pin structurally compatible to the Arduino. The block design is shown in Figure 5.

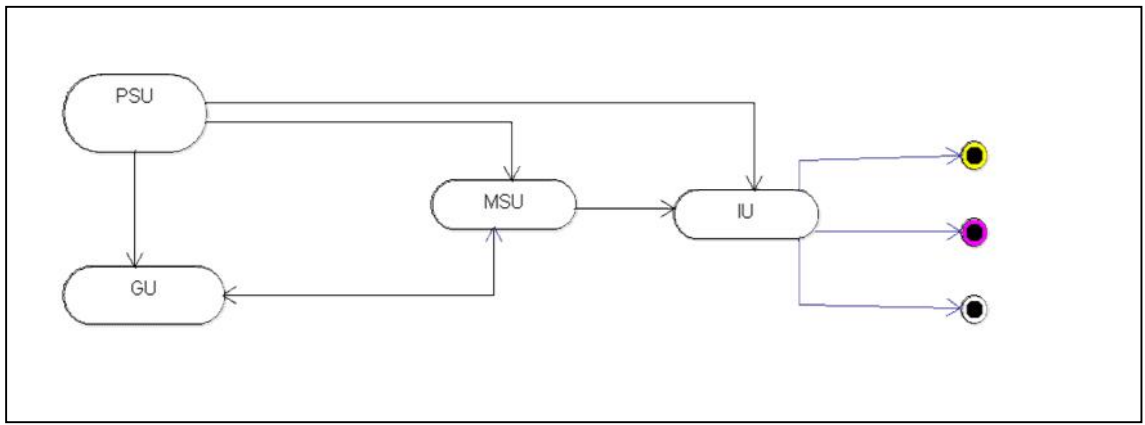

Fig 5: Systems Block Design of *GOptimaEmbed*

i. **System Components**
The initial prototype includes the power supply unit, GSM shield unit (GU), and embedded microprocessor system unit (MSU), and indications unit (IU). The embedded MSU uses the Arduino framework based on the Atmega328P RISC microcontroller/processor. IU gives status of queried species using light emitting diodes (LED's). This indication is used for real-time debugging purposes and may be excluded from the final prototype.

ii. **Specifications for Smart dbms**
The specifications include the minimum hardware specific requirements, the software and the database requirements. The algorithmic requirements shall include relevant points as described in [17]. Detailed specifications for smart microcomputer and communication module hardware can be obtained from [18] and [19] respectively. Sample data from the Iris-dataset has also been obtained from [20].

**Hardware Requirements:**
Power supply – 2A; 5v regulated power pack or 5v;6AH battery

Embedded Processor – Arduino
Current draw: 2.0A max burst
Voltage Tolerance: 7.0v-12.0v +/- 3%
GSM Module (Shield) – Quectel M10 Module
Current draw: 2.0A
Voltage Tolerance: 4.5v +/- 3%
(If using GSM module same voltage range should apply as in item II)
LED indicators – Visual Interactive Species Labels
Red – Iris-Setosa
Orange – Iris-Versicolor
Green – Iris-Virginica

**Software Requirements:**
ii. External Dependency:
iii. Class Libraries
iv. Search Engine – Genetic Algorithms

**Database requirements:**
i. Database Storage – In-memory
ii. Data Type – Integer (unsigned)
iii. Database Range : 4 – 9
iv. Number of databases – 2
v. Number of Tables – 1
vi. Number of Attributes – 3
vii. Attributes – Sepal-length, Sepal width, Petal length, Petal width
viii. Key Attribute – Sepal length

**Algorithmic Requirements:**
i. Core Input Cardinality – 4 (Platform, Population size, Number of Generations and objectives)
ii. Topology – function and procedure calls (noi. classes)
iii. Time Complexity – Logarithmic, linear
iv. Refine criterion – Worst-case
v. iTest – Turing Test

iii. **Simplifying Algorithm**

The simplified algorithm using *GOptimaEmbed* integrates the SMS-SQL translation with a Genetic Optimizer. It is possible to modify the fitness function of the optimizer to account for equivalency-to-zero checks; less-than or equal-to checks; and complex function checks either singly or in a multi-objective manner. Checks can also be made against dynamic sensor actuated inputs. Combination of various checks in various orders is also possible.

The steps taken for smart database query and report are as follows:

Step1: Initialize working variables: database store, population size, number of generations, insertion

i. IDE – Arduino 0023
and db storage variables, classes and library headers

Step2: Set Generations Counter

Step3: Initialize population

Step4: Define db_check   //Function that extracts //relevant message report from requested database // and sends to Pre-defined Client ID's.

Step5: Query (GSM)

Step6: Decode SMS strings

Step7: Translate SMS to Integer

Step8: Fitness Checks

Step9: Fit = abs(selected individual – desired //individual)//desired individual is allele in SMS //Message
Criterion: == 0; <=threshold //typical threshold = 1;
If (fitted)
Extract alleles from db storage
String/Integer match
Call (db_check)
Print Fit
Else : GoTo Step 5
Return Fit

Step10: Select Fitted Individuals

Step11: Cross-over Reproduction

Step12: Uniform Mutation

Step13: Return to Step 5

End

## 4. IMPLEMENTATION AND RESULTS

Figure 6 shows a prototype outlook of the smart system with GSM module shield interfaced to the Arduino smart computer. With the shield concept the developer can concentrate more on the implementation logic rather than on product design and thereby reducing overall system development time by half.

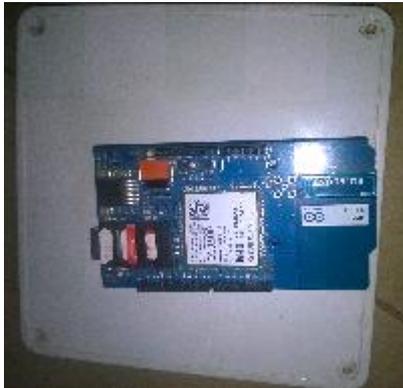

Fig 6: *GOptimaEmbed* Smart Module Interfacing

The results of simulation for various queries on the smart dbms are given in Table1. The difference between $t_{sent}$ and $t_{received}$ gives the latency which is about 2s. Users of *GOptimaEmbed* can send query using sms in the form given in section 3.3.1. The graphic report may vary depending on phone make and model.

Table 1: Query Report

| Query No. | SMS Query | $t_{sent}$ (s) | $t_{received}$ (s) | Query Report |
|---|---|---|---|---|
| 1 | dbiris tbiris atsepl va2 | 19:11p.m Feb 8 | 19:12p.m Feb 8 | Species Found Iris-Setosa |
| 2. | dbiris tbiris atsepl va2 | 19:49p.m Feb 8 | 19:50p.m Feb 8 | Species Found Iris-Setosa |
| 3. | dbiris tbiris atsepl va8 | 19:19p.m Feb 8 | 19:21p.m Feb 8 | Species Found Iris-Virginica |
| 4. | dburis tbiris atsepl va6 | 19:19p.m Feb 9 | 19:21p.m Feb 9 | Species Found IVS |

## 5. CONCLUSIONS, RECOMMENDATIONS AND FUTURE WORK

Smart dbms using *GOptimaEmbed* is a cost-effective way of implementing real-time smart solutions for information retrieval and storage in resource constrained devices. When optimized with GA's, such systems can help traditional data storage and retrieval search systems make good decisions on the desired information when faced with in-exact, fuzzy or approximate requests. Smart dbms's can also revolutionize the information engineering process and improve the fortunes of GSM operators, aid educationists and students, and anyone looking for cost-effective and ubiquitous ways of accessing their valuable information anywhere, anytime in an intelligent way.

We recommend this tool to the Nigerian Communications Commission or similar body as a useful test instrument for assessing the relative performance and quality of GSM operators in the country.

In future, we plan to integrate more SMS-SQL like functions into *GOptimaEmbed* and apply it to dynamic sensor networks for real time monitoring of plant species growth profiles.

# APPENDIX
Source-Code Listing for Implementing *GOptimaEmbed* using Arduino Framework:

```
/*

Class libraries
*/
#include <String.h>

#include <ClientGSM.h>
#include <GSM.h>
#include <inetGSM.h>
#include <LOG.h>
#include <QuectelM10.h>
#include <ServerGSM.h>
#include <Streaming.h>
#include <UDPGSM.h>
#include <WideTextFinder.h>
#include <EEPROM.h>

#include "QuectelM10.h"
#include <NewSoftSerial.h>

//Variables and Constants
const unsigned int pop_bounds[2] = {4,8};
const unsigned int pop_size = 16;
const unsigned int num_gens = 40;
int Geni;
const unsigned int num_muts = 1;
unsigned int def_length = 0;
unsigned int min_bound = 0;
unsigned int max_bound = 0;
unsigned int ipop[16];  // Initial Population
unsigned int ipopn[16]=
{5,4,7,6,7,7,4,6,5,4,7,6,7,7,4,7};  // Initial Population
// char dbase_species[8] = {'S', 'V', 'V', '71','71', '71', 'V', 'V'};
//String dbase_species[8] = {"S", "V", "V", "G","G", "G", "V", "V"};

unsigned int dbase_species[16]
={5,4,7,6,7,7,4,6,5,4,7,6,7,7,4,7};
// int dbase_species[8] = {S, 22, 22, 71, 71, 71, 22, 22};

unsigned int ipn[16];
unsigned int parent[16];
unsigned int ipop_max = 0;
unsigned int fit_pop[16];
unsigned int cfit_pop[16];
unsigned int j[16];
unsigned int fit;
unsigned int sms_in;
const unsigned int td = 5000;
unsigned int i;

unsigned int stop_criterion;

unsigned int total_fitness;

int sms_report = 0;
char smsbuffer[160];
char n[40];

unsigned int intx1 = 0;
unsigned int intx2 = 0;

//// SETTING UP
void setup() {

// randomSeed(analogRead(0));
//Light Indicators
pinMode(8,OUTPUT);
pinMode(10,OUTPUT);
pinMode(12,OUTPUT);

// Open serial communications:
Serial.begin(9600);
Serial.println("GSM Shield testing.");
//Start configuration.
if (gsm.begin())
    Serial.println("\nstatus=READY");
else Serial.println("\nstatus=IDLE");

min_bound = pop_bounds[0];
max_bound = pop_bounds[1];
def_length = (max_bound - min_bound)/pop_size;
//Serial.println("Defining Length");
    //Serial.println(def_length);

    ipop_max = pop_size*def_length + min_bound;

// send an floatro:
// sms_in = 2;
}

/// FUNCTIONS USED/CREATED

int queryGSM()
{

  gsm.readSMS(smsbuffer, 160, n, 20);

    delay(500);

}

int dbcheck_1(String int_x1, String int_x2)
{

  if(int_x1.equals("52") && int_x2.equals("57"))
```

```
    {
      switch (dbase_species[i])
      {
        case 4:
          gsm.sendSMS("07030081615","Species found is:");
          gsm.sendSMS("07030081615", "Iris-Setosa");

          gsm.sendSMS("08036710489","Species found is:");
          gsm.sendSMS("08036710489", "Iris-Setosa");

          break;
        case 5:
          gsm.sendSMS("07030081615","Species found is:");
          gsm.sendSMS("07030081615", "Iris-Setosa");

          gsm.sendSMS("08036710489","Species found is:");
          gsm.sendSMS("08036710489", "Iris-Setosa");
          break;
        case 6:
          gsm.sendSMS("07030081615","Species found is:");
          gsm.sendSMS("07030081615", "Iris-Versicolor");

          gsm.sendSMS("08036710489","Species found is:");
          gsm.sendSMS("08036710489", "Iris-Versicolor");

          break;
        case 7:
          gsm.sendSMS("07030081615","Species found is:");
          gsm.sendSMS("07030081615", "Iris-Virginica");

          gsm.sendSMS("08036710489","Species found is:");
          gsm.sendSMS("08036710489", "Iris-Virginica");
          break;
        case 8:
          gsm.sendSMS("07030081615","Species found is:");
          gsm.sendSMS("07030081615", "Iris-Virginica");

          gsm.sendSMS("08036710489","Species found is:");
          gsm.sendSMS("08036710489", "Iris-Virginica");
          break;
        //return dbcheck_1();
      }
    }
       else if(int_x1.equals("52") && int_x2.equals("69"))
    {
      switch (dbase_species[i])
      {
        case 4:
          gsm.sendSMS("07030081615", "IS");
          break;
        case 5:
          gsm.sendSMS("07030081615", "IS");
          break;
        case 6:
          gsm.sendSMS("07030081615", "IVS");
          break;
        case 7:
          gsm.sendSMS("07030081615", "IVG");
          break;
        case 8:
          gsm.sendSMS("07030081615", "IVG");
          break;
      }
    }
    delay(300);

}

/// MAIN PROGRAM STARTS HERE:
void loop()
{

//Generations Start here:
  for(Geni=0; Geni<=num_gens;Geni++)
  {

    queryGSM();

  sms_report = int(smsbuffer[23])-48;
String sepal_length = String(sms_report);
//Serial.println("sepal_length");
//Serial.println(sms_report);

sms_in = sms_report;
Serial.println("sepal_length Value = " + String(sms_in));
```

```
//Serial.println(sms_in);

intx1 = int(smsbuffer[0])-48;
String int_x1 = String(intx1);
//Serial.println("intx1");
//Serial.println(intx1);

intx2 = int(smsbuffer[2])-48;
String int_x2 = String(intx2);
//Serial.println("intx2");
//Serial.println(intx2);
delay(500);

    // GA Starts Here
   //Generation Prints Here:
   Serial.println("Generation = "+ String(Geni));
   Serial.println(Geni);

   //Serial.println("Max Population Attainable");
    //Serial.println(ipop_max);

   //Fitness Evaluation:
    for(i=0; i<=pop_size; i++)
    {
    ipop[i] = random(pop_size);//*def_length + min_bound;
     Serial.println("Population Attainable = " + String(ipop[i]));

    // fit_pop[i] = 2*def_length*ipop[i]/ipop_max;
     fit_pop[i] = abs(ipopn[ipop[i]]-sms_in);
     Serial.println("Fitted Population = " + String(fit_pop[i]));
     Serial.println("Point Locii i = " + String(i));

     //cfit_pop[i] = fit_pop[i];

  // fit_pop[i] = abs(fit_pop[i]-sms_in);
  //Serial.println("Absolute fit = ");
  //Serial.println(fit_pop[i]);

    // Stopping Criterion
   //stop_criterion = Geni;
     // if(ipopn[i] == sms_in)
     if(fit_pop[i] <= 1)
  {

   // Serial.println("cfit");
    //Serial.println(cfit_pop[i]);

    Serial.println("Solution found at Locus: " + String(i));
```

```
    Serial.println("waiting....");

   // gsm.sendSMS("07030081615","Species found is:");
    //gsm.sendSMS("07030081615", "Iris-Setosa");
    dbcheck_1(int_x1, int_x2);
    Serial.println("\nSMS sent OK");
   //break;

  delay(td);
  break;
 }

 }

//Generations and Fitness Evaluation ends here

 //Selection begins here:
 //use roulette selection (-> need pos. fitness!)

 total_fitness = fit_pop[i];
 for(i =0; i<=pop_size; i++)

 {

   ipn[i] = random(pop_size)*total_fitness;
   j[i] = abs(cfit_pop[i]-ipn[i]);
   if(j[i]== 0)

   {
    j[i] = pop_size;
     Serial.println("j[i] Empty = " + String(j[i]));
   }

   else {

    j[i] = j[1];
    Serial.println("j[i] Not Empty = " + String(j[i]));
   }

   parent[i] = ipop[j[i]];
  // parent[i] = ipopn[j[i]];
   Serial.println("parent = " + String(parent[i]));

 }
  delay(500);
  // Selection ends here
```

```
    // REPRODUCTION
// parents 2i-1 and 2i make two new children
// 2i-1 and 2i crossover
// use arithmetic crossover

for (i= 0; i<=pop_size; i= i+2)
  {

   ipop[i] = random(pop_size)+parent[i];// + (1-random(pop_size))*parent[i+1];
   ipop[i+1] = random(pop_size)+parent[i+1];

 //  ipop[i] = random(pop_size)*parent[i] + (1-random(pop_size))*parent[i+1];
  // ipop[i+1] = (1-random(pop_size))*parent[i] + random(pop_size)*parent[i+1];

  //Serial.println("Reproductive-Population");
   // Serial.println(ipop[i]);
       delay(1000);
  }
  //Cross-over script ends here

//   mutation
//   use uniform mutation
  for (i= 0; i<=num_muts; i++)
  {

ipop[random(pop_size)] = pop_bounds[0] + random(pop_size);//*def_length;

     delay(1000);
 }
 // mutation script ends here

 //;pop

 }

 delay(1000);
}
```